# NEW GENERATION OF MOBILE PHONE VIRUSES AND CORRESPONDING COUNTERMEASURES

Pu Wang[1,2,3], Marta C. González[1,3], Ronaldo Menezes[4] & Albert-László Barabási[1,5]

[1]*Center for Complex Network Research, Department of Physics, Biology and Computer Science, Northeastern University, Boston, MA 02115.*

[2] *Department of Physics, University of Notre Dame, Notre Dame, IN 46556.*

[3]*Department of Civil and Environmental Engineering, Massachusetts Institute of Technology, Cambridge, MA 02139.*

[4]*Department of Computer Sciences, Florida Institute of Technology, Melbourne, FL*

[5]*Department of Medicine, Harvard Medical School, and Center for Cancer Systems Biology, Dana Farber Cancer Institute, Boston, MA 02115.*

Email: rmenezes@cs.fit.edu

**Abstract:**

**The fast growing market for smart phones coupled with their almost continuous online presence makes these devices the new targets of virus writers. It has been recently found that the topological spread of MMS (Multimedia Message Services) viruses is highly restricted by the underlying fragmentation of the call graph. In this paper, we study MMS viruses under another type of spreading behavior: scanning. We find that hybrid MMS viruses including some level of scanning are more dangerous to the mobile community than their standard topological counterparts. However, the effectiveness of both scanning and topological behaviors in MMS viruses can generally be limited by two controlling methods: (*i*) decreasing susceptible handsets' market share (OS it runs) and (*ii*) improving monitoring capacity to limit the frequency in which MMS messages can be sent by the mobile viruses.**

# 1. Introduction:

The history of technological viruses is intrinsically linked to the history of computational devices. Since the inception of the Internet, programmers began writing self-replicating executables such as Creeper [1], the first known instance of a computer virus. From there on, the field of computer security grew together with the ability of programmers to write increasingly more sophisticated viruses. In recent years, mobile phones have become the new frontier for these self-replicating programs [2-7]. The availability of these mobile devices coupled with their continuous online presence makes them an ideal breeding ground for technological viruses [2-7].

Mobile phone viruses spread primarily through Bluetooth and MMS connections [2-7]. While a Bluetooth virus can reach the full susceptible user base, its spread is slowed by human mobility [8, 9], offering ample time for developing and deploying countermeasures [10]; Bluetooth communication has a short range and as users move away from each other the communication channel between the devices may be lost. In contrast, MMS viruses use a topological spreading approach unrestricted by user movement and can reach most susceptible users within hours of the initial attack [10]. However, their spread is limited by the market share of the OS (Operating System) they run which leads to a fragmentation of the underlying call graph [11, 12] making it hard for a virus to cover the entire graph. In this paper, we study MMS viruses under the assumption that viruses are able to scan for numbers in the call graph; what we call *scanning spreading approach*.

Scanning is a simple form of viral behavior. It implies the use of Pseudo-Random Number Generators (PRNGs) to select a target within the population of handsets. Once the target number is selected, the virus sends a copy of itself to that target via MMS. The mobile-phone world has seen instances of this behavior such as the Timofonica virus [13] and more recently in a hybrid virus called Beselo [14]. At the first glance, random scans seem to be a naïve approach but what makes them dangerous is that a few successful scans may transfer the infection from one side of the network to another in a very short amount of time. Figure 1 depicts the approaches of topology vs. scanning. Note that part of the network (depicted in light yellow) could never be reached without scans. This example demonstrates the difference between topological viruses (which have to respect the existing connections between users) and scanning viruses (which are able to jump from user to user in the network). Below the phase transition point [15-20] of OS market share, no giant component exists in the network [10, 15-20], however, scanning links change the network topology, making the giant component exist even market share is under the phase transition point.

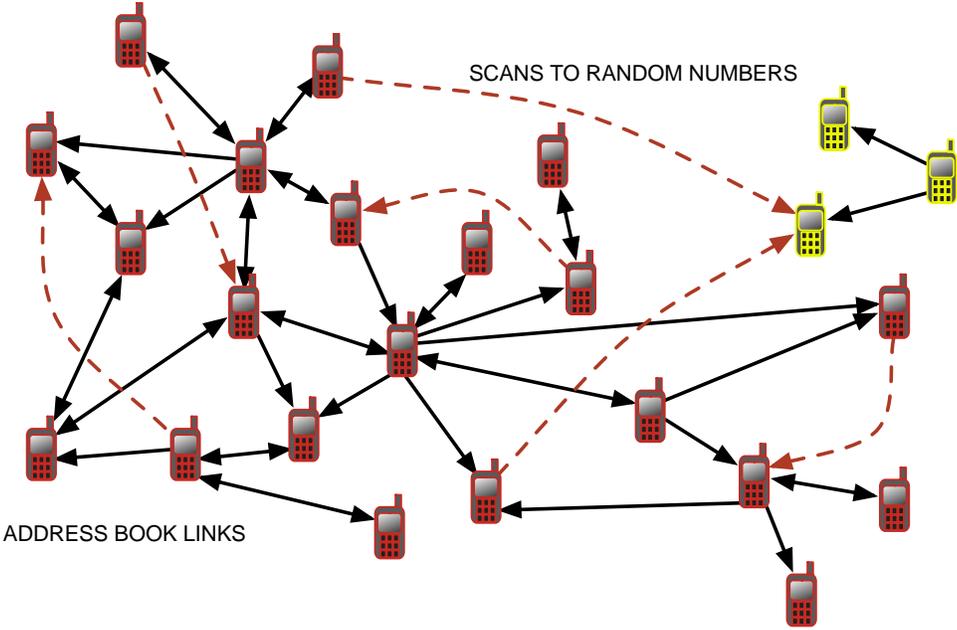

**Figure 1.** We can observe that scans are independent of the topology of the network. In other words, the topology formed by users having other users' numbers in their phone books is not used for the scans (in dashed line).

## 2. Dataset and Methodology:

### 2.1 MMS usage pattern

The dataset used in this paper was collected by a mobile phone operator for billing and operational purposes during 12 weeks period. The privacy of all callers is ensured through the use of a security key (hash code) for each user instead of users' real phone numbers.

We start our study with an analysis of the use of MMS as form of communication. Figure 2 shows the result of such analysis for the MMS activity of approximately 6 million mobile-phone users over a period of 12 weeks with an average volume of 4.7 million messages per week. The figure shows periodic usages of MMS peaking from Sunday to Tuesday. This pattern is widely used by phone operators to protect communication system by monitoring abnormal MMS activities [5, 12]. Moreover, in our simulations we assume that mobile phone operators are able to use the global activity patterns to check for anomalies that may arise from big fluctuations in individuals' MMS activity. In the insert of Figure 2, we measure the maximum and average MMS volume in different two-hour periods for a week (Figure 2a). If the MMS volume generated by the spread of viruses is larger than the volume difference, $\Delta V$, between the maximum and average MMS volume (Figure 2b), phone operators are able to detect the viruses using simple anomaly detection approaches, otherwise, MMS viruses may spread without being detected (phone operators generally regard these slightly higher rates of messages as part of expected fluctuations in users' MMS usage [12]).

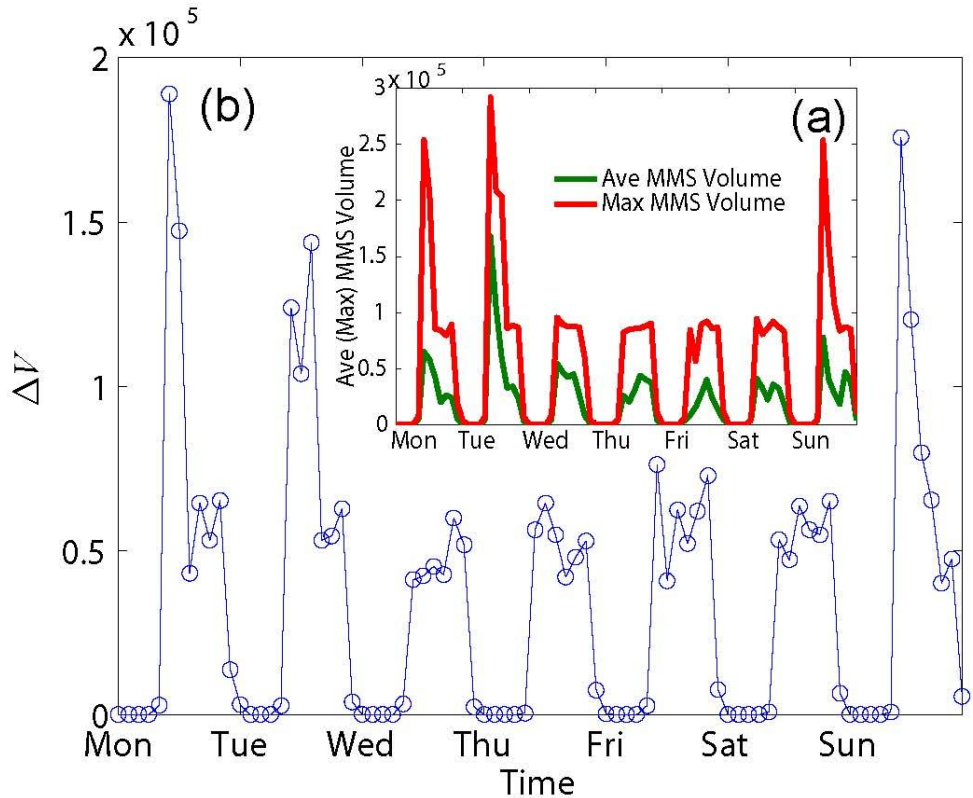

**Figure 2.** The insert (a) shows the maximum and average MMS volumes calculated from 12 weeks' mobile phone data. The difference between the two volumes, $\Delta V$, is shown in (b) and is used as a threshold of MMS viruses being detected by phone operators.

**2.2 *SI* Model and the topological and scanning approaches of MMS viruses**

We use the *SI* model [21] to simulate the spread of MMS viruses. Under this model, mobile phones can be in only one of two possible states: infected (*I*), when they are transmitting the infection, or susceptible (*S*), when they are vulnerable to infections. In our simulations we are interested in the initial spreading process in the absence of recovery or antiviral software, so we do not consider the possibility that the phones could recover from the infection, a reasonable assumption due to the limited capacity of some handsets for installing antiviral software [3-5], combined with the users' current lack of concern about the threat of mobile phone viruses [3-5]. Although many mobile phone viruses require user confirmation to install (disguised as a photo or a music file) [3-5], in our simulations we assume that the virus does not need a user confirmation, corresponding to the worst possible spreading scenario.

In a standard *SI* model an infected mobile phone can infect a susceptible phone at a rate $\mu$ described by

$$S + I \xrightarrow{\mu} 2I \tag{1}$$

Thus the number of infected users (*I*) evolves in time as

$$\frac{dI}{dt} = \beta SI / N \tag{2}$$

We define the effective infection rate as $\beta=\mu<k>$, where $\mu=1/\tau$ is the virus's infection rate, or the inverse of $\tau$ (the time it takes for the virus to infect a susceptible handset) multiplied by $<k>$ which represents the average number of contacts in a typical user's address book. In this paper, we study two spreading approaches of MMS viruses: in a topological spread the infected handset sends out malicious MMS to the phone numbers listed in its address book; and scanning spread where the handset sends out malicious MMS to random generated phone numbers.

In the topological approach, we approximate a user's address book with the list of numbers the user communicated with during one month of observation. In the simulation, we assign the initial virus to a randomly chosen handset in the network, which in turn will send infected MMSs to all its identified contacts. The MMS service is not instantaneous – there is some time delay on receiving a MMS. In our simulation we choose $\tau=2$ minutes as the time required for a MMS virus to be received by another handset and to install itself [22]. We choose each simulation time step as 2 minutes, and set $\mu=1$ and $<k>=1$. That is, once infected a mobile phone may infect all of its contacts.

For the scanning approach, we define the effective scanning probability $p$ as the probability for scans reaching active phone numbers. To estimate this probability, we divided 6 million (an approximation of the mobile phone user base) by 100 million (the total phone numbers that 8 digits can generate), obtaining an effective scanning probability $p=0.06$. We further postulate a higher probability $p=0.25$ to study the scenario that viruses improve their scanning ability to reach more active phone numbers.

To quantify the topological/scanning approach level in the spread of MMS viruses, we define the random attack probability, $\rho$, that a virus will attack a random phone number rather than a number listed in the address book. Its value varies from zero to one, representing viruses' different attack strategies. For example, $\rho=0$ means pure topological attack strategy and $\rho=1$ means pure scanning attack strategy.

Viruses generally limit and control their attack times to prevent them being detected from the abnormal MMS volume they caused. Hence, we define the maximum attack number $s$ for each infected handset, which is the total number of viral MMSs that an infected phone can send.

## 2.3 The naïve and temporal spread model

In this paper, we study two spreading models: *naïve* and *temporal*. The difference between them is on their ability of utilizing temporal patterns of MMS volume to prevent being detected. With the information about MMS volume during a day, the temporal spread model aims at understanding viruses that try to avoid detection by phone providers. On the other hand, the naïve model is studied to understand the worst-case scenario of a viral outbreak, ignoring MMS temporal usage patterns and possible monitoring by phone operators. Without loss of generality, we use OS market share values of $m=0.30$ and $m=0.03$ to study the viral spread in different types of call graphs [11, 12]. When the market share $m=0.30$ there is a giant component in the call graph and in the small $m=0.03$ case no giant component exists and call graph is fragmented into small isolated clusters.

## 3. Results and Analysis

### 3.1 The effect of scanning

We start by looking at the effect of topological and random attacks on a real call graph. We study a small neighborhood of the call graph (~600 nodes) generated by starting from a randomly chosen user and including all mobile phone contacts up distance 4 from it. The nodes in Figure 3a are represented using two colors that correspond to the two kinds of OSs with 25% and 75% market share respectively. A virus infects only the OS it was designed for, thus the largest components [15-20] formed by the same color connected nodes represent the maximum number of handsets that a virus can infect. Without random scans, the red nodes distribute in small islands and the largest component size represents only 6% of the total number of red nodes (see Figure 3b). If random scanning links are added (see the dashed lines in Figure 3c), one can see that the structure of the call graph changes significantly, the largest component size is about 6 times bigger, revealing that the addition of scanning links makes the virus more dangerous.

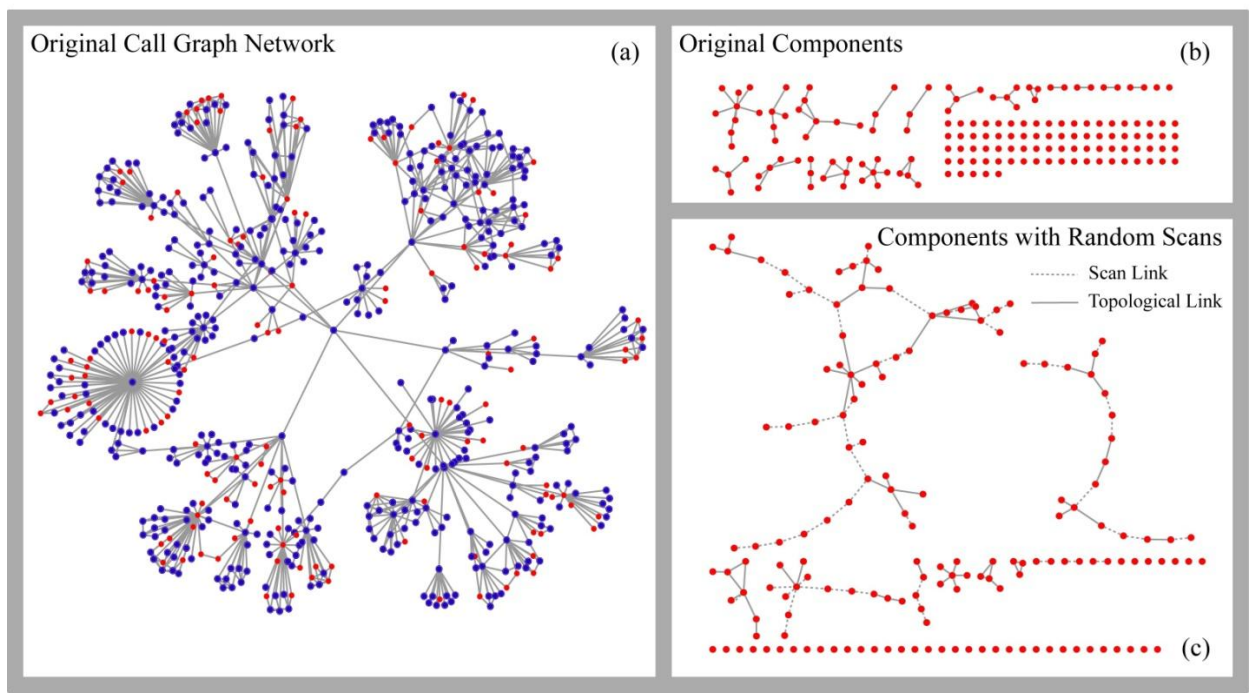

**Figure 3.** (a) A small neighborhood of the call graph is generated by starting from a randomly chosen mobile phone user and including all mobile phone contacts up distance 4 from it. The two colors used in the nodes represent two kinds of OSs with 25% and 75% market share respectively. (b) Without the random scans, the red nodes distribute in small islands; the largest component size represents only 6% of the total number of red nodes. (c) With random scans, small isolated clusters are connected, forming larger clusters (the dash lines represent the connections created by random scans); the size of the largest component increases to 35% of the total number of red nodes.

## 3.2 Naïve Viral Spread Model (Worst Case Study)

In the naïve viral spread model, the viral spread is added uniformly on top of the normal daily MMS activity. This model studies the worst-case of a virus spread, since it looks at the epidemics under the assumption that no protection or detection mechanisms are in place. We measured the infection fraction ($I/N$, the total number of infected users over the total susceptible user base) for different epidemic scenarios of $m$ (market share), $s$ (maximum attack number), $p$ (effective-scanning probability) and $\rho$ (random-attack probability). For each case, we run ten independent simulations with the virus starting at randomly chosen users.

Intuitively, one can see the different characteristics of topological and scanning approaches. Topological attacks always reach active phone numbers but they are generally trapped in isolated clusters of the underlying fragmented call graph. In contrast, scanning attacks have a much lower possibility to reach active phone numbers, but just a few of them can link the isolated clusters together. In Figures 4a & 4b we find that for a small market share $m=0.03$, a relatively big random attack probability $\rho$ can connect the fragmented call graph, making MMS viruses infect more susceptible handsets. This can be explained by scanning attacks' ability to connect isolated clusters. However, pure scanning attacks ($\rho=1$) may result in the failure of spread due to its low probability to reach active phone numbers. For a large market share $m=0.3$, interestingly, the most threatening attack strategy is related to the maximum attack number $s$. As depicted in Figure 4c, when the maximum attack number $s$ is small, MMS viruses with more topological attacks will cause more damages because scanning attacks have a low ability to reach active phone numbers. If the maximum attack number $s$ is big ($s=50$), we find that MMS viruses with a random attack probability $\rho\sim0.7$ are the most dangerous, because topological attacks will be ineffective if repeating to attack same infected phones iteratively. Hence, we conclude here that some attack strategies can increase the danger posed by MMS viruses.

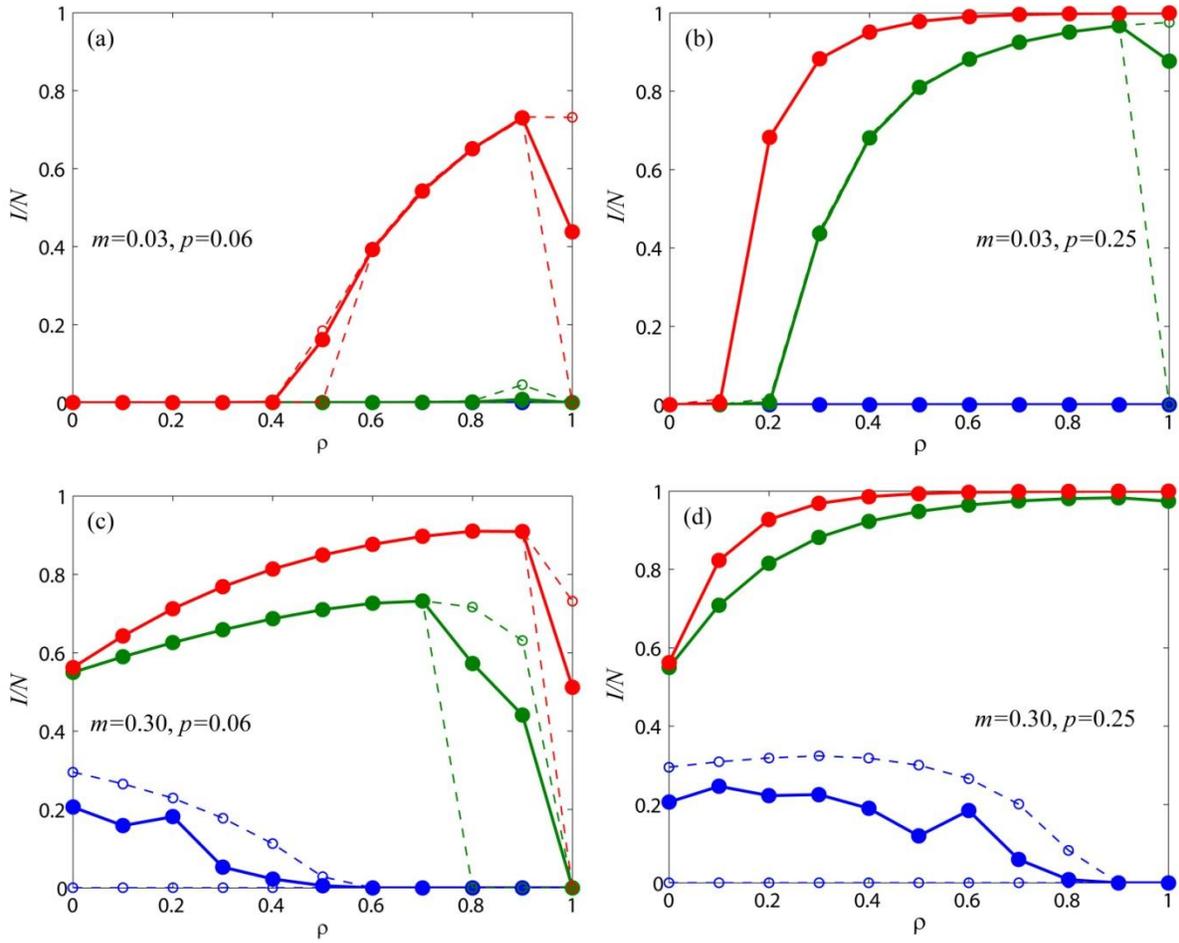

**Figure 4.** The spread using the naïve viral spread model. The solid and dashed lines represent the average (solid line), maximum (dashed line) and minimum (dashed line) infection fraction (*I/N*) respectively. Parameter *m* is the market share and parameter *p* is the effective scan probability. Different colors of lines and symbols correspond to different maximum attack numbers *s*=100 (blue), *s*=500 (green), *s*=1000 (red) for market share *m*=0.03; and *s*=10 (blue), *s*=50 (green), *s*=100 (red) for market share *m*=0.3. All of the results are from 10 simulations where the virus starts at a randomly chosen mobile-phone user.

Following the approach in [10], we study how the market share *m* influences the spread of MMS viruses. We set maximum attack number *s* at different values, from 10 to 1000. For a small market share *m*=0.03, independently of what attack strategies (*p* value) the viruses utilize, they cannot spread if *s* is smaller than 100. In contrast, for a large market share *m*=0.30, there is a giant component formed even when the maximum attack number *s* is as small as 10. Therefore, large market share handsets are much more vulnerable than their smaller counterparts. This strengthens the results in [10], revealing again the crucial role of OS market share in the spread of mobile phone viruses.

Next, we explore the function of maximum attack number *s* in the spread of MMS viruses. In Figures 4a and 4b, as we increase this number from 100 to 1000, the virus can infect more than 70% of the susceptible user base. Similar results are found in Figures 4c and 4d, the infection fraction (*I/N*) grows quickly as *s* increases from 10 to 50. Hence, besides market share *m*, the maximum attack number *s* also plays a key role in the spreading process. This finding provides us an effective way to defend MMS viruses: if we successfully constrain viruses' maximum attack number *s*, our communication system can be safe independent of what attack strategies viruses use.

We also find that the larger the scan effective probability *p* is, the more danger the virus poses. However, comparing with the functions of market share and maximum attack number in the spread, its contribution is not so prominent.

From the analysis above, we discover the conditions that lead to the successful spread of MMS viruses. On the other hand, we find crucial elements to stop or prevent a viral outbreak. At time of writing, we are not familiar with a virus strategy based on the value of *p* (random attack probability). However, based on our findings we can force viruses' attack strategies to fail. Phone operators and anti-viral software companies could decrease susceptible phones' market share *m* by adding protections to more phones. They could also improve their monitoring capacity, detecting MMS viruses earlier and forcing viruses to limit their maximum attack number *s*. The limitation of viruses' maximum attack number can significantly slow down the spread of MMS viruses, driving them to vanish after a few iterations (infections).

The results above provide an indication of the values for market share as well as maximum attack number in which the mobile-phone base becomes susceptible to global epidemics. Unfortunately however this is not the entire story as viruses are being written to be more stealth to detection. In the next section we delve into a spread mode that attempts to be stealth by using the patterns in MMS volume on different times of the day and different days of the week.

### 3.2 Temporal Volume-based Viral Spread Mode (The Danger of Being Stealth)

Given that the pattern of MMS virus can be analyzed, a MMS virus may utilize latent periods of usage to avoid detection by the phone operators. One of the common ways virus can do the above is by using an approach based on the time of the day and day of the week, as well as limiting each infected phone's maximum attack number. In the model proposed here, MMS viruses spread solely during the daytime according to the temporal MMS volume pattern and with each infected phone's maximum attack number *s* and average attack period *T* being set. For example, if 2% of the weekly MMS volume is generated between 6:00pm to 8:00pm on Tuesday and each infected phone sends out one viral MMS on average per day (*T*=1 day), we calculate the probability of the infected phone sending out a viral message in each 2 minutes [22] simulation step as following:

(1) Since the average attack period is *T*=1 day, an infected handset would send out seven viral MMSs per week.

(2) Between 6:00pm to 8:00pm on Tuesday, two percent of the total weekly MMS volume is generated, thus the infected handset has a probability of 7*0.02=0.14 to send out a viral MMS between 6:00pm to 8:00pm on Tuesday.

(3) Between 6:00pm to 8:00pm, there are sixty two-minutes time steps, thus we get that the infected handset has a probability of $7 \times 0.02/60 = 0.0023$ to send out a viral MMS in each two-minutes during 6:00pm to 8:00pm.

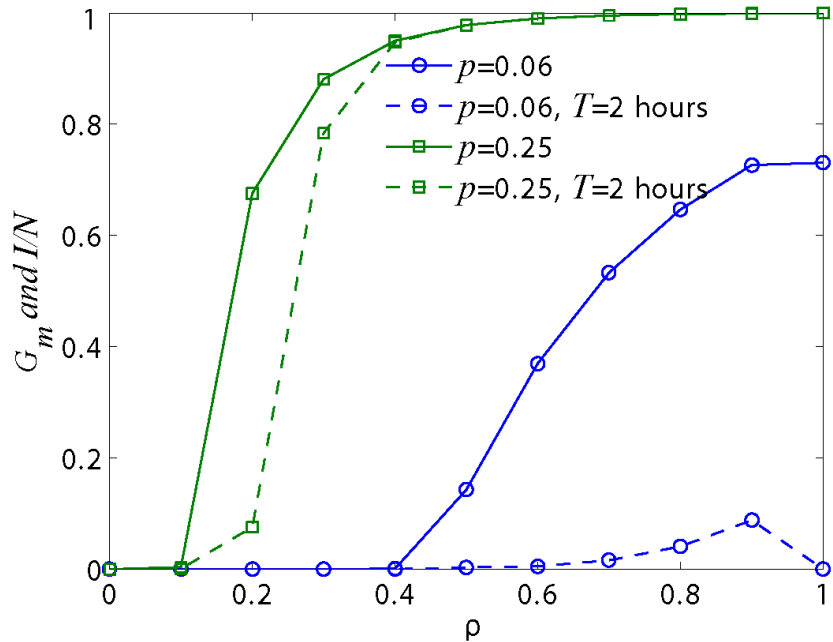

**Figure 5.** The final ratio of infected and susceptible handsets (*I/N*) in one year observational period and the giant component size $G_m$ in *m*=0.03 case. As we set the MMS viruses' attack period as *T*=2 hours, the ratio of infected handsets is usually smaller than the giant component size $G_m$ due to the slow spreading mechanism.

First we study the spread under different average attack periods *T*. That is, we calculate the ratio of infected and susceptible handsets (*I/N*) for different average attack periods *T*, finding that large attack period (low attack frequency) can make the final infection ratios smaller than the giant component size $G_m$ (the maximum number of susceptible phones that the virus can infect). This not only tells us that low attack frequencies will decrease the final infection ratio and even stop the virus at its initial stage, but again reminds us of the important countermeasure to protect the communication system: improve our monitoring capacity to detect the virus early or force the virus to use a low-attack frequency.

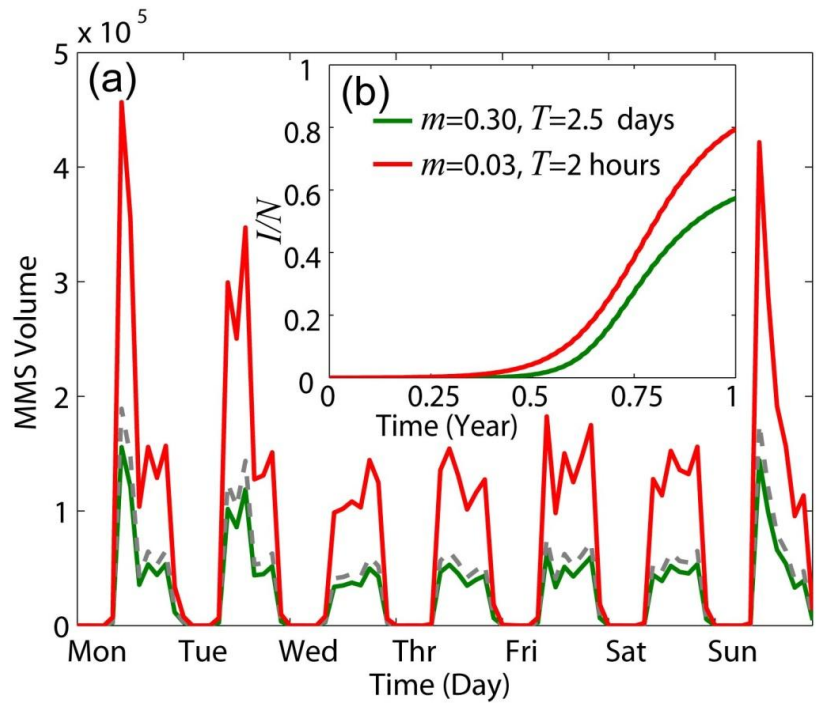

**Figure 6.** The MMS volume created by the temporal viral spread of MMS viruses where the red, green and dashed lines correspond respectively to the results with market share *m*=0.03, *m*=0.30 and the detection threshold. The inset shows the epidemic curves for market share *m*=0.03 and *m*=0.30 cases.

An important question we would like to answer is whether MMS viruses could infect most susceptible users without being detected by the phone operator. It is obvious that the smaller the average attack period, the faster the virus spreads but that would make it very easily detectable. For large market shares such as $m=0.30$, we perform the following experiments, we first set the parameters $s$(maximum attack number)=50, $\rho$(random-attack probability)=0.1, $p$(effective-scanning probability)=0.25 and $T$(average attack period)=2.5 day. We find that the virus can infect about one million users in about one year and the MMS volume created by the viruses is well below the threshold volume of being detected. That means that given enough time it can infect a large fraction of susceptible phones without being found. For a small market share $m=0.03$, we set the parameters $s$(maximum attack number)=1000, $\rho$(random attack probability)=0.3, $p$(effect scanning probability)=0.25 and $T$(average attack period)=2 hour. We find the virus can infect 0.14 million susceptible users in one year and the MMS volume resulted by the viruses exceeds the threshold of being detected (see Figure 6). This is because the small market-share of susceptible phones results in the low efficiency of attacks.

From the results above, we see that a virus can infect a large number of users in a stealth mode. They can later make the infected phones perform some simple malicious functions, such as sending short text messages to get the communication channels jammed – a typical denial-of-service attack. On the other hand, we demonstrate the effectiveness of anomaly detection schemes in detecting the virus or forcing the virus to spread very slowly, providing the phone operators with a guide to put in place proper countermeasures.

## 4. Conclusions and Discussions:

We studied the interplay between topological and scanning behaviors of MMS viruses. For a small market share with no giant component existing in the call graph, the addition of random scans can significantly increase the possibility of an epidemic outbreak. For a large market share with a giant component in the call graph, the addition of random scans can only slowly increase the number of infected phones or even drive the virus to stop itself. We also find that given enough time, some viruses can infect a large fraction of susceptible phones without being detected by phone operators. Fortunately, independent of MMS viruses' attack strategies, the epidemics can still be limited by the OS market share and smart anomaly detection schemes. Added to a good understanding of the network formed from connections between users, smart anomaly detection schemes may be able to prevent mobile phones to become the next platform for virus writers hence avoiding the situation typical in computer systems where virus writers seem to be winning the battle.

**Acknowledgements:**

We thank G. Xiao and C. Song for discussions and comments on the manuscript. This work was supported by the James S. McDonnell Foundation 21st Century Initiative in Studying Complex Systems, the National Science Foundation within the DDDAS (CNS-0540348), ITR (DMR-0426737) and IIS-0513650 programs, and the US Office of Naval Research Award N00014-07-C.